# Proton and Neutron Pairing Properties within a mixed volume-surface pairing force using SKI3-HFB Theory

Malik A. Hasan[*1] and Ali H. Taqi[1]

[1]*Department of Physics, College of Science, University of Kirkuk, Kirkuk, Iraq.*

**ABSTRACT**

This work aims at a systematic investigations of the pairing properties and Fermi properties from the proton drip-line to the neutron drip-line. In order to provide more accurate mass formula with skyrme SKI3 force, the global descriptive power of the SKI3-HFB model for pairing properties are made in this study. Systematic Skyrme SKI3-Hartree-Fock-Bogoliubov calculations with a mixed volume-surface pairing force are carried out to study the ground-state neutron and proton pairing gap, neutron and proton pairing energy and neutron and proton Fermi energy for about 2095 even-even nuclei ranging from $2 \leq Z \leq 110$ to $2 \leq N \leq 236$. The calculated results of neutron and proton pairing gap are compared with experimental data using the difference-point formulas $\Delta^{(3)}, \Delta^{(4)}$ and $\Delta^{(5)}$, and also compared with the neutron and proton pairing gap of Lipkin-Nogami model. It is shown that the Skyrme-SKI3 functional with the mixed volume-surface pairing force can be successfully used for describing the ground-state pairing and Fermi properties of the investigated nuclei, in particularly the neutron-rich nuclei and the exotic nuclei near the neutron drip-line. On the other hand, the calculated neutron and proton pairing gap are in good agreement with the available experimental values of the neutron and proton pairing gap of the difference-point formulas $\Delta^{(3)}, \Delta^{(4)}$ and $\Delta^{(5)}$ and with the data of Lipkin-Nogami model over the whole nuclear chart.

## 1. INTRODUCTION

The pairing correlations can be considered as a major ingredient in describing the nuclear ground-state properties for the finite nuclei. It is based on the so-called BCS approximation, which was introduced by Bardeen-Cooper-Schrieffer in superconductivity theory in metals. The nuclear pairing correlations were established for the first time since 60 years [1, 2]. Till today, the study of the structure of exotic nuclei which are lying away from the beta-stability line and close to the proton or neutron drip-line play an important role in describing and understanding the low-lying nuclear ground-state properties. Those nuclei (exotic or halo nuclei) which are lying in the regions far from the $\beta$-stability line has attracted wide attention and represented one of the most open and active areas of research in the field of nuclear physics, both theoretically and experimentally [3-7].

To describe the nuclear ground-state properties such as pairing gap and pairing energy of the exotic nuclei, especially on those regions with neutron-rich side of the beta-stability valley, an appropriate theoretical framework is needed to describe these exotic regions [8, 9]. The Hartree-Fock-Bogoliubov (HFB) approach is a good theoretical framework to study and describe the nuclear pairing correlations [10-16]. In general, the HFB theory is consisting of two parts: first, the self-consistent mean-field from the Hartree-Fock (HF) theory which describes the long-range part as the particle-hole (*p-h*) channel used in closed-shell configurations, and the second part is the pairing correlations obtained from the Bardeen-Cooper-Schrieffer (BCS) theory [17] of superconductivity in metals, which describes the short-range part as the particle-particle (*p-p*) channel used in open-shell configurations. By depending on the two channels, the density matrix ($\rho_{ij}$) and the pairing tensor density ($\kappa_{ij}$) can be used to characterize the nuclear system. The two parts were unified and achieved in a general HFB formalism using a variational principle [18, 19].

In the previous paper [20], we have studied the ground-state binding energy, two-neutron separation energy ($S_{2n}$), quadrupole deformation parameter ($\beta_2$), charge radii, neutron and proton rms radii and neutron pairing gap of the whole nuclear chart. Our previous study [20] with the Skyrme-SKI3 force showed success in describing the ground-state properties by comparing the calculated results with the experimental data and other models results such as Finite Range Droplet Model (FRDM), Relativistic Mean-Field (RMF) model and HFB calculations based on D1S Gogny force. The authors *S. A. Changizi, et al.* in Refs. [21, 22] presented a systematic study on the neutron pairing gaps predicted by HFB calculations with the Skyrme-SLy4 parameter with different volume, surface, and mixed volume-surface pairing interactions. Thus, in order to make a comprehensive perception to describe the nuclear landscape with different Skyrme functionals, here, in the present work, we have examined and investigated the neutron and proton pairing properties and neutron and proton Fermi properties for about 2095 even-even nuclei with 2 (He) $\leq Z \leq$ 110 (Ds) over a wide range of isotopes starting from the neutron number $N = 2\ to\ N = 236$ ($4 \leq A \leq 346$). The investigated properties of the whole nuclear chart are based on the HFB theory with Skyrme-SKI3 parameter [23] with a mixed volume-surface pairing force. The calculated results of neutron and proton pairing gap have been compared with the neutron and proton pairing gap of the available experimental data of difference-point formulas $\Delta^{(3)}, \Delta^{(4)}$ and $\Delta^{(5)}$, and also compared with the neutron and proton pairing gap of Lipkin-Nogami model ($\Delta_{n,p}^{LN}$) data [24-26] to show the consistence of the calculated results and the validity of our model. It is necessary to mentioned that the studies of the whole nuclear chart with the Skyrme-SKI3 parameter are few somewhat in the nuclear structure, for this reason; this force (SKI3) was chosen in this study, because it gives a good predictions and results in describing the nuclear structure (see Ref. [20] for example).

The structure of this paper is organized as follows. In Sec. 2 a brief theoretical framework of the HFB approach is presented. The numerical details of this study are given in Sec. 3. The results and discussion are given in Sec. 4. Finally, the conclusions are presented in Sec. 5.

## 2. THE HFB FORMALISM

The HFB approach has been extensively discussed in the literature [10-12], and we will only be briefly introduced here for simplicity. In the standard HFB formalism, a two-body Hamiltonian of a system of fermions can be expressed in terms of a set of annihilation and creation operators $(c, c^\dagger)$ [10, 11]:

$$H = \sum_{n_1 n_2} e_{n_1 n_2} c^\dagger_{n_1} c_{n_2} + \frac{1}{4} \sum_{n_1 n_2 n_3 n_4} \bar{v}_{n_1 n_2 n_3 n_4} c^\dagger_{n_1} c^\dagger_{n_2} c_{n_4} c_{n_3} \qquad (1)$$

Where the first term corresponding to the kinetic energy and $\bar{v}_{n_1 n_2 n_3 n_4} = \langle n_1 n_2 | V | n_3 n_4 - n_4 n_3 \rangle$ is the matrix element of the two-body interaction between anti-symmetrized two-particle states. The HFB ground-state wave function $|\Phi\rangle$ is defined as the quasiparticle vacuum $\alpha_k |\Phi\rangle = 0$, where the quasiparticle operators $(\alpha, \alpha^\dagger)$ are connected to the original particle operators via a linear Bogoliubov transformation [10, 11]:

$$\alpha_k = \sum_n (U^*_{nk} c_n + V^*_{nk} c^\dagger_n), \quad \alpha^\dagger_k = \sum_n (V_{nk} c_n + U_{nk} c^\dagger_n) \qquad (2)$$

which can be rewritten in the matrix form as:

$$\begin{pmatrix} \alpha \\ \alpha^\dagger \end{pmatrix} = \begin{pmatrix} U^\dagger & V^\dagger \\ V^T & U^T \end{pmatrix} \begin{pmatrix} c \\ c^\dagger \end{pmatrix} \qquad (3)$$

The matrices $U$ and $V$ satisfy the relations [10, 11]:

$$U^\dagger U + V^\dagger V = I \qquad \qquad U U^\dagger + V^* V^T = I$$
$$\text{and} \qquad (4)$$
$$U^T V + V^T U = 0 \qquad \qquad U V^\dagger + V^* U^T = 0$$

In term of the normal density $\rho$ and pairing tensor $\kappa$, the density matrices of one-body are:

$$\rho_{nn'} = \langle \Phi | c_{n'}^\dagger c_n | \Phi \rangle = (V^* V^T)_{nn'}, \quad \kappa_{nn'} = \langle \Phi | c_n c_n | \Phi \rangle = (V^* U^T)_{nn'} \quad (5)$$

The expectation value of the Eq. (1) can be expressed in an energy functional as [10, 11]:

$$E[\rho, \kappa] = \frac{\langle \Phi | H | \Phi \rangle}{\langle \Phi | \Phi \rangle} = \mathrm{Tr}\left[\left(e + \frac{1}{2}\Gamma\right)\rho\right] - \frac{1}{2}\mathrm{Tr}[\Delta \kappa^*] \quad (6)$$

where the self-consistent term is:

$$\Gamma_{n_1 n_3} = \sum_{n_2 n_4} \bar{v}_{n_1 n_2 n_3 n_4} \rho_{n_4 n_2} \quad (7)$$

and the pairing field term is:

$$\Delta_{n_1 n_2} = \frac{1}{2} \sum_{n_3 n_4} \bar{v}_{n_1 n_2 n_3 n_4} \kappa_{n_3 n_4} \quad (8)$$

The variation of the energy (Eq. (6)) with respect to $\rho$ and $\kappa$ leads to the HFB equation [10, 11]:

$$\begin{pmatrix} e + \Gamma - \lambda & \Delta \\ -\Delta^* & -(e+\Gamma)^* + \lambda \end{pmatrix} \begin{pmatrix} U \\ V \end{pmatrix} = E \begin{pmatrix} U \\ V \end{pmatrix} \quad (9)$$

where the Lagrange multiplier $\lambda$ has been introduced to fix the correct average particle number, and $\Delta$ denote the pairing potential.

## 3. NUMERICAL DETAILS

In this paper, the ground-state properties of the even-even nuclei have been investigated by using the code HFBTHO (v2.00d) [12] with Skyrme-SKI3 functional [23], which utilizes the axial Transformed Harmonic Oscillator (THO) single-particle basis to expand quasi-particle wave functions. It iteratively diagonalizes the HFB Hamiltonian based on generalized Skyrme-like energy densities and a zero-range pairing interaction until a self-consistent solution is found [11,12].

The calculations were performed with a mixed volume-surface pairing force has been used with a cutoff quasi-particle energy $E_{cut} = 60$ MeV, which means that all the quasiparticles with energy lower than the cutoff are taken into account in the calculations of the densities. The Harmonic Oscillator (HO) basis was characterized by the oscillator length $b = -1.0$ fm which means that the code automatically sets $b_0$ by using the relation of HO frequency [12]:

$$b_0 = \sqrt{\hbar/m\omega}, \text{ with } \hbar\omega = 1.2 \times 41/A^{1/3} \tag{10}$$

To obtain more accurate results, the number of oscillator shells (the principal number of oscillator shells $N$) taken into account was $N_{\max}=20$ shells. The axial deformation parameter ($\beta_2$) of the basis taken into account was 0.2 (the deformation value 0.2 is a value that is mediate the values of large and small deformation). The input data of the pairing strength in the code HFBTHO (v2.00d) in Eq. (11) for neutrons $V_0^n$ and protons $V_0^p$ have been used as a pre-defined pairing force depending on a standard value of each Skyrme functional used in the code HFBTHO. In the case of the SKI3 force, the pairing strength for neutrons $V_0^n = -357.23$ MeV and for protons $V_0^p = -388.56$ MeV. It is always assumed that the pairing force reads:

$$V_{pair}^{n,p}(\mathbf{r}) = V_0^{n,p}\left(1 - \alpha\frac{\rho(\mathbf{r})}{\rho_c}\right)\delta(\mathbf{r} - \mathbf{r}') \tag{11}$$

where $\rho(\mathbf{r})$ is the local density, and $\rho_c$ is the saturation density, fixed at $\rho_c = 0.16$ fm$^{-3}$, and the type of pairing force is defined by the parameter $\alpha$, which can be volume character, surface, or mixed volume-surface characteristics [12]. More about numrical detailes can be found in Refs. [20, 27].

## 4. RESULTS AND DISCUSSION

In this section, the ground-state properties of the pairing profile and Fermi properties for about 2095 even-even nuclei with $2\,(He) \leq Z \leq 110\,(Ds)$ over a wide range of isotopes starting from the neutron number $N = 2$ to $N = 236$ ($4 \geq A \leq 346$) will be investigated and discussed by using the HFB-Skyrme SKI3 force method with the mixed volume-surface zero-range pairing force interaction.

### 4.1 Neutron and Proton Pairing Gap

In Fig. 1(left panel), the calculated results of 759 even-even neutron pairing gap from $Z=2$, $N=4$ to $Z=110$, $N=170$ with Skyrme-SKI3 force have been plotted as a function of the proton ($Z$) and neutron ($N$) number, and compared with the available experimental data of many finite-difference formulas, which are often interpreted as a measurement of the empirical pairing gap such as the three-point $\Delta_{(Z)}^{(3)}(N)$ formula [28, 29]:

$$\Delta_{(Z)}^{(3)}(N) \equiv \frac{\pi_N}{2}[BE(Z, N-1) + BE(Z, N+1) - 2BE(Z, N)] \tag{12}$$

And the four $\Delta_{(Z)}^{(4)}(N)$ and five-point formula $\Delta_{(Z)}^{(5)}(N)$ can be defined as [28, 30-33]:

$$\Delta_{(Z)}^{(4)}(N) \equiv \frac{\pi_N}{4}[BE(Z, N-2) + 3BE(Z, N) - 3BE(Z, N-1) - BE(Z, N+1)] \tag{13}$$

$$\Delta_{(Z)}^{(5)}(N) \equiv -\frac{\pi_N}{8}[BE(Z, N+2) + 6BE(Z, N) + BE(Z, N-2)$$

$$-4BE(Z, N-1) - 4BE(Z, N+1)] \tag{14}$$

where $\pi_N = (-1)^N$ represents the number of parity and the experimental values are calculated from the binding energy (*BE*) given in Ref. [34] using the three-point formula $\Delta_{(Z)}^{(3)}(N)$, four-point formula $\Delta_{(Z)}^{(4)}(N)$ and five-point formula $\Delta_{(Z)}^{(5)}(N)$, as seen in Fig. 1(right panel).

In Fig. 1(left panel), the calculated results of neutron pairing gap are ranged from 0 MeV for many nuclei (for example $^{16}$O) up to the maximum value $\cong$2.96 MeV at *Z*=2, *N*=4 ($^6$He), whereas the value of $\Delta_{(Z)}^{(3)}(N)$ are ranged between 0.53 MeV at *Z*=104, *N*=160 ($^{264}$Rf) up to the maximum value 6.88 MeV at *Z*=6, *N*=6 ($^{12}$C), the value of $\Delta_{(Z)}^{(4)}(N)$ are ranged between 0.46 MeV at *Z*=104, *N*=160 ($^{264}$Rf) up to the maximum value 6.36 MeV at *Z*=4, *N*=4 ($^8$Be), and the value of $\Delta_{(Z)}^{(5)}(N)$ are ranged between 0.52 MeV at *Z*=104, *N*=160 ($^{264}$Rf) up to the maximum value 5.97 MeV at *Z*=4, *N*=4 ($^8$Be) as seen in Fig. 1(right panel).

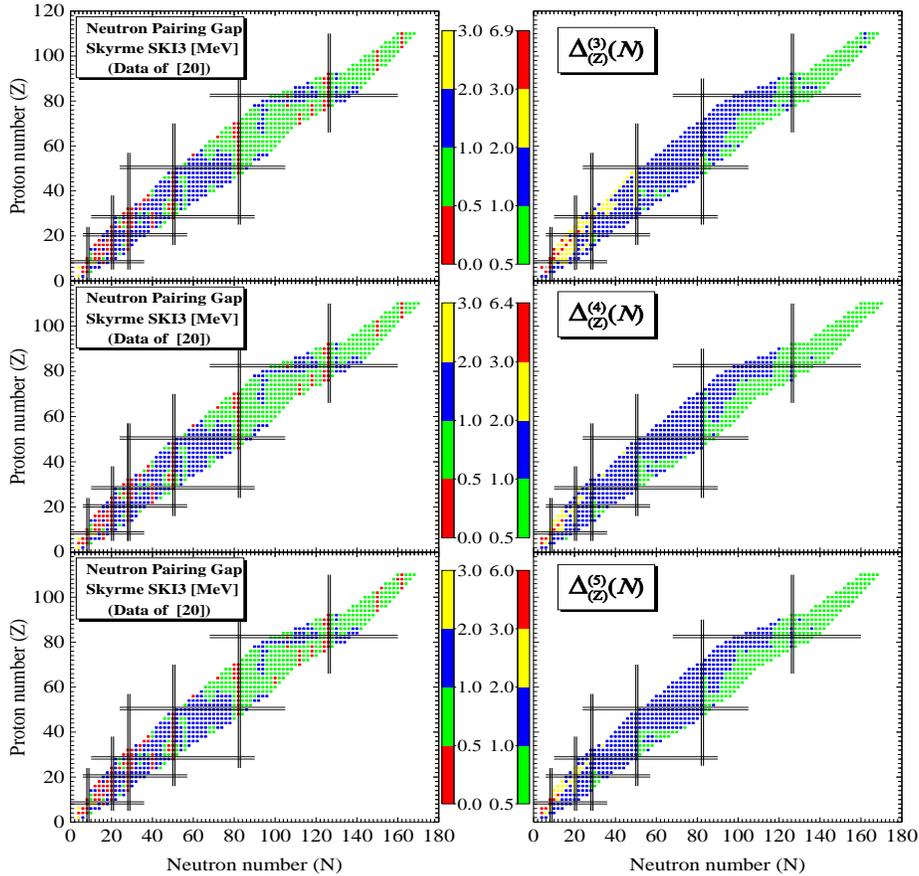

FIG. 1 (Color online) HFBTHO calculations with the mixed volume-surface pairing force for 759 even-even nuclei as a function of proton ($Z$) and neutron ($N$) number: (left panel) the calculated neutron pairing gap with Skyrme SKI3 force, in comparison with the $\Delta^{(3)}_{(Z)}(N)$, $\Delta^{(4)}_{(Z)}(N)$ and $\Delta^{(5)}_{(Z)}(N)$ values (right panel).

Figure 2(left panel) shows the calculated neutron pairing gap in comparison with the $\Delta^{(3)}_{(Z)}(N)$, $\Delta^{(4)}_{(Z)}(N)$ and $\Delta^{(5)}_{(Z)}(N)$ values as a function of neutron number ($N$). The values of the calculated results with Skyrme SKI3 force started with a high values at the light nuclei regions (from $Z=2$, $N=4$ to $Z=20$, $N=24$) and decreased as the number of neutrons increases. In general, the calculated results shows clear consistency and good agreement with the values of $\Delta^{(3)}_{(Z)}(N)$, $\Delta^{(4)}_{(Z)}(N)$ and $\Delta^{(5)}_{(Z)}(N)$ formulas. The same behavior can be seen in Fig. 2(right panel) which shows the calculated neutron pairing gap in comparison with the $\Delta^{(3)}_{(Z)}(N)$, $\Delta^{(4)}_{(Z)}(N)$ and $\Delta^{(5)}_{(Z)}(N)$ values as a function of proton number ($Z$). It is worth to mention that the values of the calculated neutron pairing gap for most nuclei that have magic (or semi-magic) numbers with $Z$, $N=8$, 20, 28, 50, 82, and 126 found to be with a zero values, while the difference pairing gap formulas ($\Delta^{(3)}_{(Z)}(N)$, $\Delta^{(4)}_{(Z)}(N)$ and $\Delta^{(5)}_{(Z)}(N)$) does not have a zero values for the nuclei with magic (or semi-magic) numbers.

As depicted in Fig. 2(left panel), the values of both of the calculated neutron pairing gap and $\Delta^{(3)}_{(Z)}(N)$, $\Delta^{(4)}_{(Z)}(N)$ and $\Delta^{(5)}_{(Z)}(N)$ formulas of the neutron magic numbers ($N=50$, 82 and 126) are shown as a separated regions (or discrete regions) from the other neighboring nuclei regions, and the regions of those nuclei which located between the magic numbers have somewhat a large values of neutron pairing gap. In contrast to the Fig. 2(left panel), these regions ($N=50$, 82 and 126) appear as peaks (or regions somewhat with high values) as shown in Fig. 2(right panel), and the regions of those nuclei which located between the magic numbers have somewhat a low values of neutron pairing gap for both of the calculated neutron pairing gap and $\Delta^{(3)}_{(Z)}(N)$, $\Delta^{(4)}_{(Z)}(N)$ and $\Delta^{(5)}_{(Z)}(N)$ formulas.

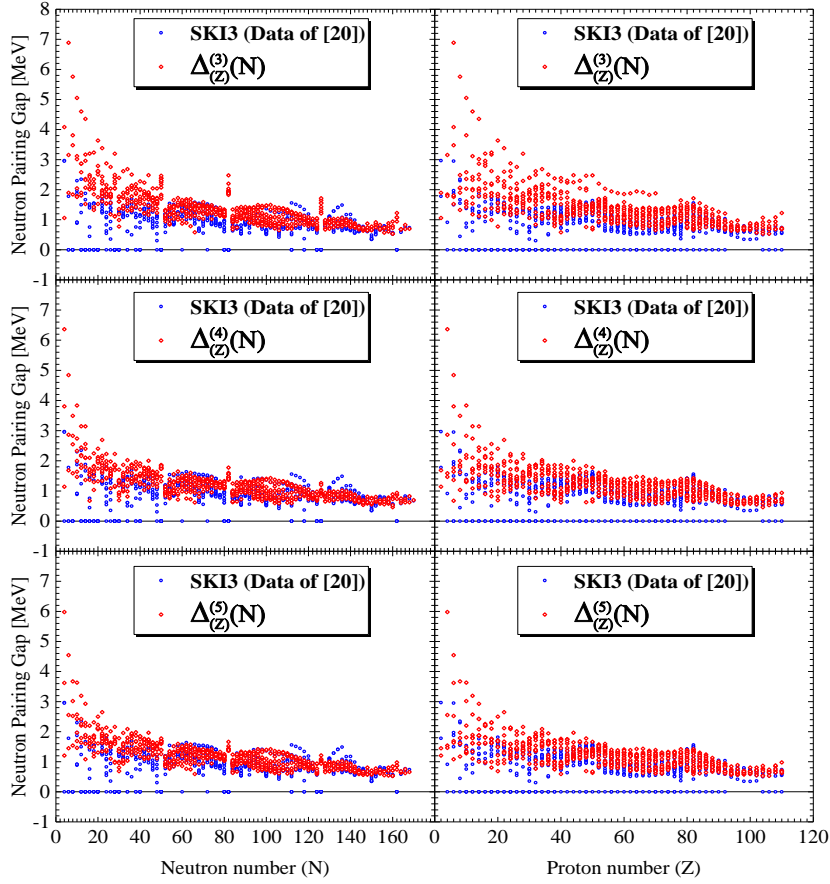

FIG. 2 (Color online) Skyrme SKI3-HFBTHO calculations of neutron pairing gap with the mixed volume-surface pairing force for 759 even-even nuclei in comparison with the $\Delta^{(3)}_{(Z)}(N)$, $\Delta^{(4)}_{(Z)}(N)$ and $\Delta^{(5)}_{(Z)}(N)$ values: (left panel) as a function of neutron number (*N*) and (right panel) as a function of proton number (*Z*).

The proton pairing gap can be expressed in the same way as in the neutron pairing gap, only by keeping the neutron number (*N*) constant an varying the proton number (*Z*) around neighboring isotones. In Fig. 3(left panel), the calculated results of 759 even-even proton pairing gap from *Z*=4, *N*=2 to *Z*=110, *N*=170 with Skyrme-SKI3 force have been plotted as a function of the proton (*Z*) and neutron (*N*) number, and compared with the available data of many finite-difference formula such as the three-point $\Delta^{(3)}_{(N)}(Z)$ formula [28, 29]:

$$\Delta^{(3)}_{(N)}(Z) \equiv \frac{\pi_Z}{2}[BE(N, Z-1) + BE(N, Z+1) - 2BE(Z, N)] \quad (15)$$

And the four $\Delta^{(4)}_{(N)}(Z)$ and five-point formula $\Delta^{(5)}_{(N)}(Z)$ can be defined as [28, 30-33]:

$$\Delta^{(4)}_{(N)}(Z) \equiv \frac{\pi_Z}{4}[BE(N, Z-2) + 3BE(Z, N) - 3BE(N, Z-1) - BE(N, Z+1)] \quad (16)$$

$$\Delta_{(N)}^{(5)}(Z) \equiv -\frac{\pi_Z}{8}[BE(N, Z+2) + 6BE(Z, N) + BE(N, Z-2) - 4BE(N, Z-1) - 4BE(N, Z+1)] \tag{17}$$

In Fig. 3(left panel), the calculated results of proton pairing gap are ranged from 0 MeV for many nuclei (for example $^{40}$Ca) up to the maximum value $\cong 3.44$ MeV at $Z=4$, $N=6$ ($^{10}$Be), whereas the value of $\Delta_{(N)}^{(3)}(Z)$ are ranged between 0.748 MeV at $Z=104$, $N=150$ ($^{254}$Rf) up to the maximum value 8.72 MeV at $Z=4$, $N=4$ ($^{6}$Be), the value of $\Delta_{(N)}^{(4)}(Z)$ are ranged between 0.53 MeV at $Z=40$, $N=66$ ($^{106}$Zr) up to the maximum value 6.18 MeV at $Z=4$, $N=4$ ($^{8}$Be), and the value of $\Delta_{(N)}^{(5)}(Z)$ are ranged between 0.63 MeV at $Z=104$, $N=152$ ($^{256}$Rf) up to the maximum value 5.97 MeV at $Z=4$, $N=4$ ($^{8}$Be) as seen in Fig. 3(right panel).

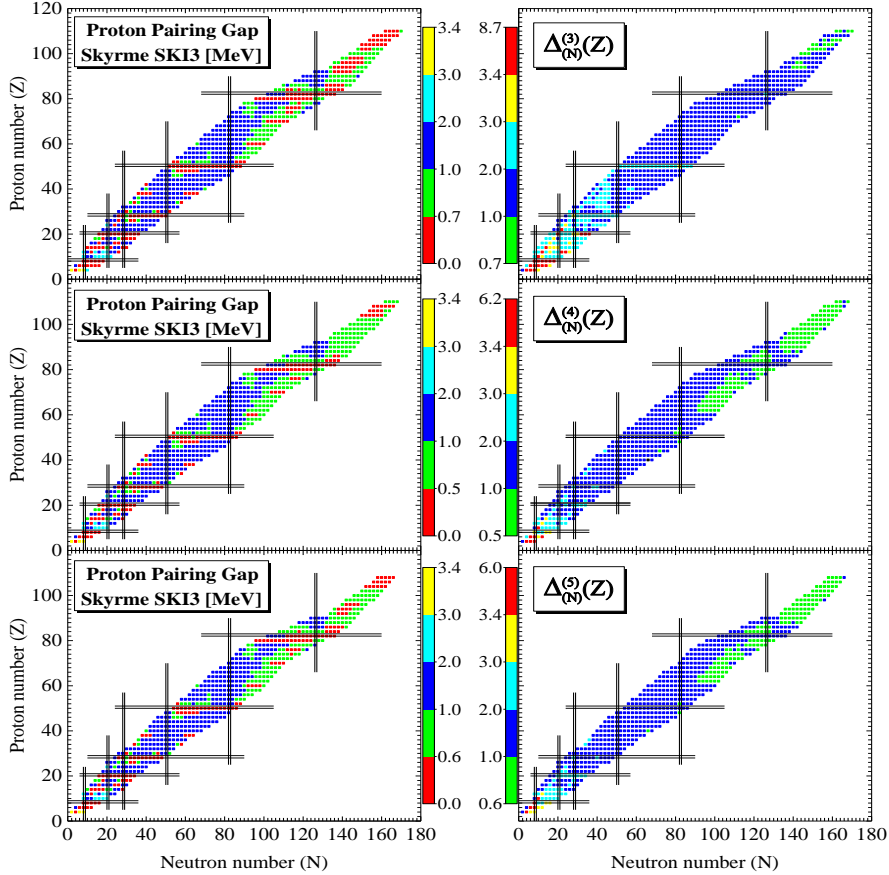

FIG. 3 (Color online) HFBTHO calculations with the mixed volume-surface pairing force for 759 even-even nuclei as a function of proton ($Z$) and neutron ($N$) number: (left panel) the calculated proton pairing gap with Skyrme SKI3 force, in comparison with the $\Delta_{(N)}^{(3)}(Z)$, $\Delta_{(N)}^{(4)}(Z)$ and $\Delta_{(N)}^{(5)}(Z)$ values (right panel).

Figure 4(left panel) shows the calculated proton pairing gap in comparison with the $\Delta^{(3)}_{(N)}(Z)$, $\Delta^{(4)}_{(N)}(Z)$ and $\Delta^{(5)}_{(N)}(Z)$ values as a function of neutron number ($N$). The values of the calculated results with Skyrme SKI3 force started with a high values at the light nuclei regions (from $Z=2$, $N=4$ to $Z=20$, $N=24$) and decreased as the number of neutrons increases. In general, the calculated results shows clear consistency with the values of $\Delta^{(3)}_{(N)}(Z)$, $\Delta^{(4)}_{(N)}(Z)$ and $\Delta^{(5)}_{(N)}(Z)$ formulas (especially with $\Delta^{(4)}_{(Z)}(N)$ formula). The same behavior can be seen in Fig. 4(right panel) which shows the calculated proton pairing gap in comparison with the $\Delta^{(3)}_{(N)}(Z)$, $\Delta^{(4)}_{(N)}(Z)$ and $\Delta^{(5)}_{(N)}(Z)$ values as a function of proton number ($Z$). It is worth to mention that the values of the calculated proton pairing gap for most nuclei that have magic (or semi-magic) numbers with $Z$, $N=8$, 20, 28, 50, 82, and 126 found to be with a zero values, while the difference pairing gap formulas ($\Delta^{(3)}_{(N)}(Z)$, $\Delta^{(4)}_{(N)}(Z)$ and $\Delta^{(5)}_{(N)}(Z)$) does not have a zero values for the nuclei with magic (or semi-magic) numbers.

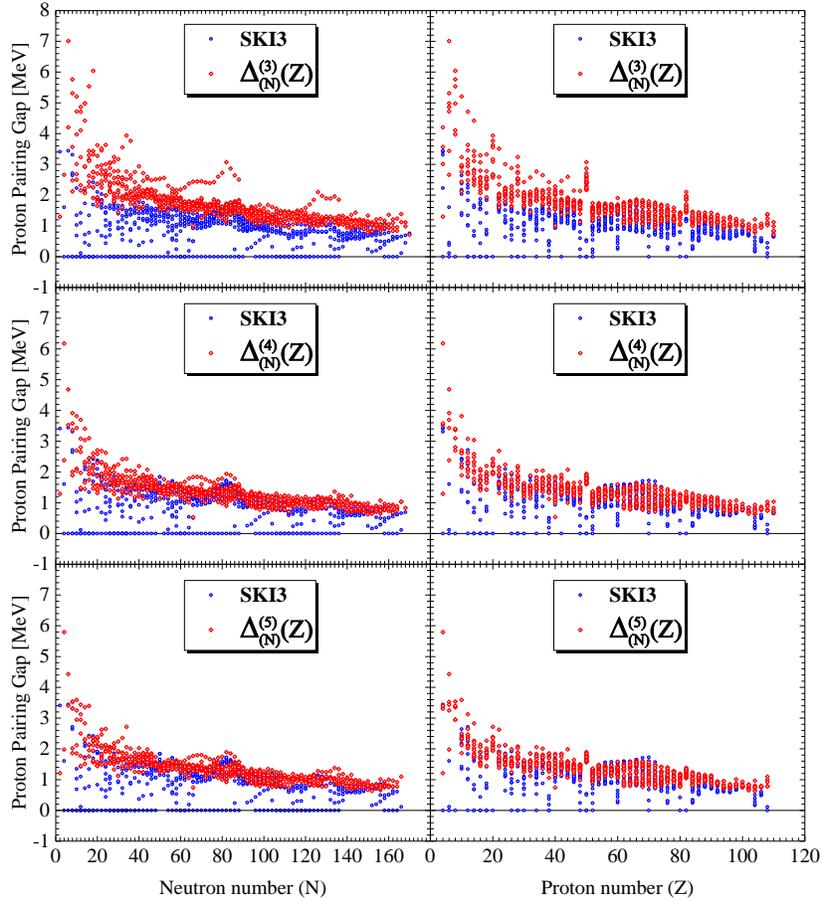

FIG. 4 (Color online) Skyrme SKI3-HFBTHO calculations of proton pairing gap with the mixed volume-surface pairing force for 759 even-even nuclei in comparison with the $\Delta^{(3)}_{(N)}(Z)$, $\Delta^{(4)}_{(N)}(Z)$ and $\Delta^{(5)}_{(N)}(Z)$

values: (left panel) as a function of neutron number (*N*) and (right panel) as a function of proton number (*Z*).

In Fig. 5(a, c), the calculated results of even-even neutron ($\Delta_n^{SKI3}$) and proton $\Delta_p^{SKI3}$) pairing gap with Skyrme-SKI3 force have been plotted as a function of the proton (*Z*) and neutron (*N*) number, respectively, and compared with the available data of the neutron ($\Delta_n^{LN}$) and proton ($\Delta_p^{LN}$) pairing gap of Lipkin-Nogami model [24] as seen in Fig. 5(b, d), respectively. In Fig. 5(a), the calculated values of $\Delta_n^{SKI3}$ for 2076 even-even nuclei from *Z*=8, *N*=8 ($^{16}$O) to *Z*=110, *N*=236 ($^{346}$Ds) are ranged from 0 MeV for many nuclei (for example $^{16}$O) up to the maximum value ≅2.34 MeV at *Z*=8, *N*=12 ($^{20}$O), whereas the values of $\Delta_n^{LN}$ are ranged between 0.5 MeV at *Z*=106, *N*=232 ($^{338}$Sg) up to the maximum value 3.66 MeV at *Z*=8, *N*=8 ($^{16}$O) as seen in Fig. 5(b). The values of $\Delta_n^{SKI3}$ with $8 \leq Z, N \leq 50$ show that this regions is not so regular as in the same region of $\Delta_n^{LN}$ values. However, the regions above $Z, N > 50$ (green color with 0.5 MeV ≤ *neutron pairing gap* ≤ 1 MeV) up to the end of the chart shows clear consistency with the values of $\Delta_n^{LN}$, except for some small regions ($N \geq 82$) with yellow color (0 → 0.5 MeV) and red color (1.5 → 2 MeV). Moreover, the large values of $\Delta_n^{SKI3}$ and $\Delta_n^{LN}$ are concentrated in light nuclei regions ($N \leq 35$) as clear in Fig. 5(a, b) for $\Delta_n^{SKI3}$ and $\Delta_n^{LN}$, respectively.

In Fig. 5(c), the calculated/measured values of $\Delta_p^{SKI3}$ for 2076 even-even nuclei from *Z*=8, *N*=8 ($^{16}$O) to *Z*=110, *N*=236 ($^{346}$Ds) are ranged from 0 MeV for many nuclei (for example $^{56}$Ni) up to the maximum value ≅2.72 MeV at *Z*=12, *N*=8 ($^{20}$Mg), whereas the values of $\Delta_p^{LN}$ are ranged between 0.72 MeV at *Z*=108, *N*=160 ($^{268}$Hs) up to the maximum value 3.58 MeV at *Z*=8, *N*=10 ($^{18}$O) as seen in Fig. 5(d). The same behavior as in Fig. 5(a) appears for the values of $\Delta_p^{SKI3}$ with $8 \leq Z, N \leq 50$ is that this regions is not so regular as in the same region of $\Delta_p^{LN}$ values. However the regions above $Z, N > 50$ (green and blue colors with 0.72 and 1 MeV ≤ *proton pairing gap* ≤ 1 and 1.5 MeV, respectively) up to the end of the chart shows clear consistency with the values of $\Delta_p^{LN}$, except for some small regions ($N \geq 82$) with red color (1.5 ↔ 2 MeV) that appear in $\Delta_p^{SKI3}$ chart and does not found in $\Delta_p^{LN}$ chart. Moreover, the large values of $\Delta_p^{SKI3}$ and $\Delta_p^{LN}$ are concentrated in light nuclei regions ($N \leq 43$) as clear in Fig. 5(c, d) for $\Delta_p^{SKI3}$ and $\Delta_p^{LN}$, respectively. Another note that can be seen in Fig. 5(c) is that the values of the $\Delta_p^{SKI3}$ for all the nuclei with the double magic or semi-magic numbers (*Z, N*=2, 8, 20, ..., 184) are 0 → 0.72 MeV as expected. This behavior is not found in proton pairing gap with Lipkin-Nogami model $\Delta_p^{LN}$.

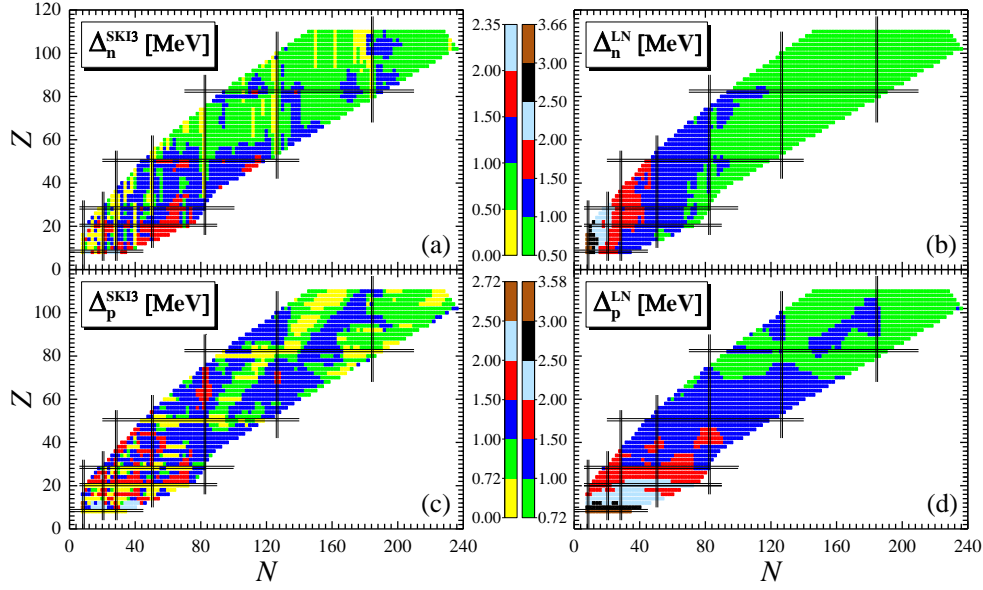

FIG. 5 (Color online) HFBTHO calculations with the mixed volume-surface pairing force for 2076 even-even nuclei as a function of proton ($Z$) and neutron ($N$) number: (a) and (c) calculated neutron and proton pairing gap with Skyrme SKI3 force, respectively, in comparison with (b) and (d) neutron and proton pairing gap with Lipkin-Nogami model [24], respectively.

In Fig. 6(a, c), The calculated results of $\Delta_n^{SKI3}$ and $\Delta_p^{SKI3}$ with Skyrme-SKI3 force have been plotted as a function of the neutron ($N$) number, respectively, and compared with the available data of $\Delta_n^{LN}$ and $\Delta_p^{LN}$ with Lipkin-Nogami model [24] as shown in Fig. 6(b, d), respectively. The values of $\Delta_n^{SKI3}$ and $\Delta_p^{SKI3}$ started with a large values in light nuclei regions and begin to decreases toward the neutron drip-line as the neutron number increases for both the calculated results $\Delta_n^{SKI3}$, $\Delta_p^{SKI3}$ and the Lipkin-Nogami $\Delta_n^{LN}$, $\Delta_p^{LN}$ data. The calculated results of $\Delta_n^{SKI3}$ and $\Delta_p^{SKI3}$ are not so regular as in Lipkin-Nogami nuclei $\Delta_n^{LN}$ and $\Delta_p^{LN}$, for neutron and proton, respectively. As depicted in Fig. 6(a), the $\Delta_n^{SKI3}$ of nuclei with the most neutron magic numbers ($N$=82, 126 and 184) are shown as a separated regions (or discrete regions) from the other neighboring regions. It should be mentioned that most the nuclei with the neutron magic numbers have a $\Delta_n^{SKI3}$ value close to the zero line (i. e. between 0 → 0.7 MeV). This behavior appears in Fig. 6(b) as regions of low values of $\Delta_n^{LN}$. In contrast to the Fig. 6(a, b), these regions ($N$=82, 126 and 184) appear as peaks as shown in Fig. 6(c, d) for neutron and proton pairing gap, respectively. Remarkably, the regions of those nuclei which located between the magic numbers have a low values of $\Delta_p^{SKI3}$ and $\Delta_p^{LN}$ which clearly seen as fluctuations in Fig. 6(c, d).

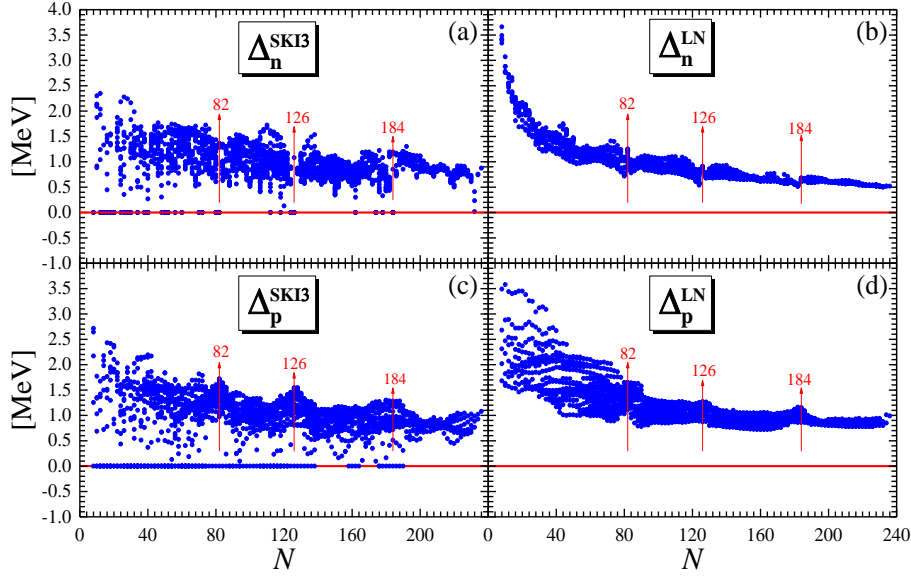

FIG. 6 (Color online) Same as Fig. 5 but as a function of the neutron number *N*.

## 4.2 Neutron and Proton Pairing Energy and Fermi Energy

The pairing energy and the Fermi energy are an important features in description the nuclear properties. From the Fermi energy we can obtain an important information especially of the exotic nuclei near/at the neutron drip-line region. It is showing the boundaries of the nuclear landscape [25, 26].

The calculated results of neutron pairing energy $E_\Delta^n$ [Fig. 7(a)], proton pairing energy $E_\Delta^p$ [Fig. 7(b)], neutron Fermi energy $\lambda_n$ [Fig. 7(c)] and proton Fermi energy $\lambda_p$ [Fig. 7(d)] using mixed volume-surface pairing interaction with Skyrme-SKI3 force have been plotted as a function of the proton (*Z*) and neutron (*N*) number. In Fig. 7(a), the calculated values of $E_\Delta^n$ for 2095 even-even nuclei from $Z=2$, $N=2$ ($^4$He) to $Z=110$, $N=236$ ($^{346}$Ds) are ranged from 0 MeV for many nuclei (for example $^4$He) up to the maximum value $\cong -29.888$ MeV at $Z=46$, $N=110$ ($^{156}$Pd). The values of $E_\Delta^p$ are ranged between 0 MeV at $Z=2$, $N=2$ ($^4$He) up to the maximum value $-21.992$ MeV at $Z=100$, $N=126$ ($^{226}$Fm) as seen in Fig. 7(b). Whereas the values of $\lambda_n$ are ranged between $-30.504$ MeV at $Z=6$, $N=2$ ($^8$C) up to the maximum value 2.96 MeV at $Z=20$, $N=72$ ($^{92}$Ca) as seen in Fig. 7(c) and the values of $\lambda_p$ are ranged between $-36.522$ MeV at $Z=8$, $N=34$ ($^{42}$O) up to the maximum value 3.889 MeV at $Z=52$, $N=48$ ($^{100}$Te) as seen in Fig. 7(d)

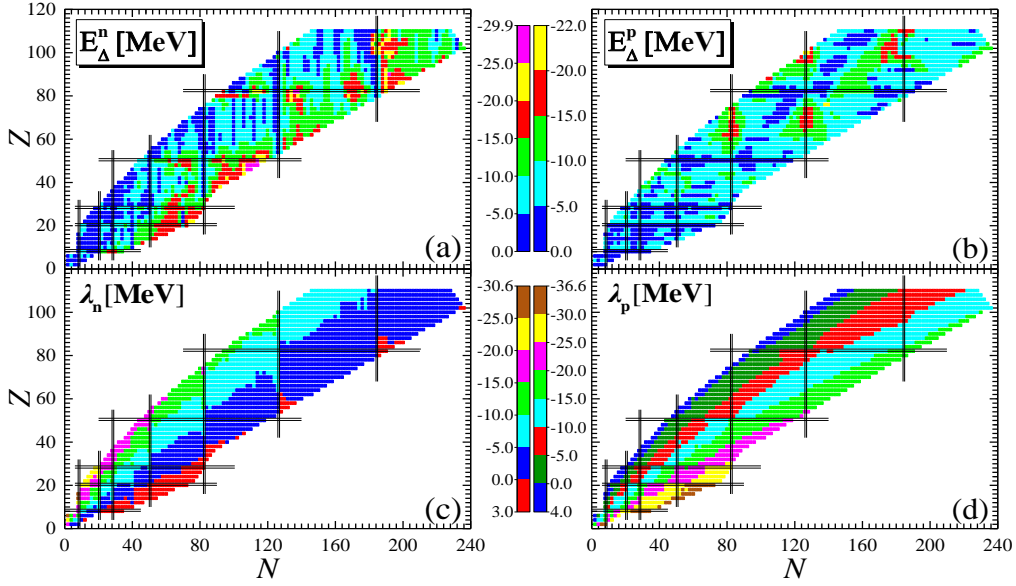

FIG. 7. (Color online) HFBTHO calculations with the mixed volume-surface pairing force interaction for 2095 even-even nuclei as a function of (Z) and (N): (a) and (b) calculated neutron and proton pairing energy with Skyrme SKI3 force, respectively, (c) and (d) calculated neutron and proton Fermi energy with Skyrme SKI3 force, respectively.

The calculated results of $E_\Delta^n$, $E_\Delta^p$, $\lambda_n$ and $\lambda_p$ with Skyrme-SKI3 force have been plotted as a function of the neutron (N) number as shown in Fig. 8(a, b, c, d), respectively. As can be seen from Fig. 8(a) that the regions between the magic numbers (N=50, 82, 126 and 184) have a large values of neutron pairing energy $E_\Delta^n$, and appear as a peaks. It should be mentioned that most the nuclei with the neutron magic numbers have a $E_\Delta^n$ value close to the zero line (i. e. between $0 \to -0.5$ MeV). In contrast to the Fig. 7(a), the regions with the magic numbers (N=50, 82, 126 and 184) have a large values of proton pairing energy $E_\Delta^p$ and appear as a peaks as shown in Fig. 8(b). Remarkably, the regions of those nuclei which located between the magic numbers have a low values of $E_\Delta^p$ which clearly seen as fluctuations in Fig. 8(b). The neutron Fermi energy $\lambda_n$ in Fig. 8(c) start with large values at the light regions of nuclei, then as the neutron number begin to increase, the $\lambda_n$ begin to decrease towards the neutron drip-line until reached the zero line.

In the regions with the magic numbers (N=50, 82 and 126), we notice that the $\lambda_n$ values for these numbers are higher than (sharp drop) the neighbouring regions, and this property can be observed, which appears in Fig. 8(c), as prominent peaks. The positive energy values of $\lambda_n$ are concentrated in the nuclei regions with: $12 \leq N \leq 94, 128 \leq N \leq 136$ and $186 \leq N \leq 200$, whereas the positive energy values of $\lambda_p$ are focused in the nuclei regions with $2 \leq N \leq 154$ as can be seen in Fig. 8(c, d) for neutron and proton

Fermi energy, respectively. The $\lambda_p$ values of the isotopic chain are more regular in the process of decreasing with the increasing of the neutron number *N*, as shown in Fig. 8(d).

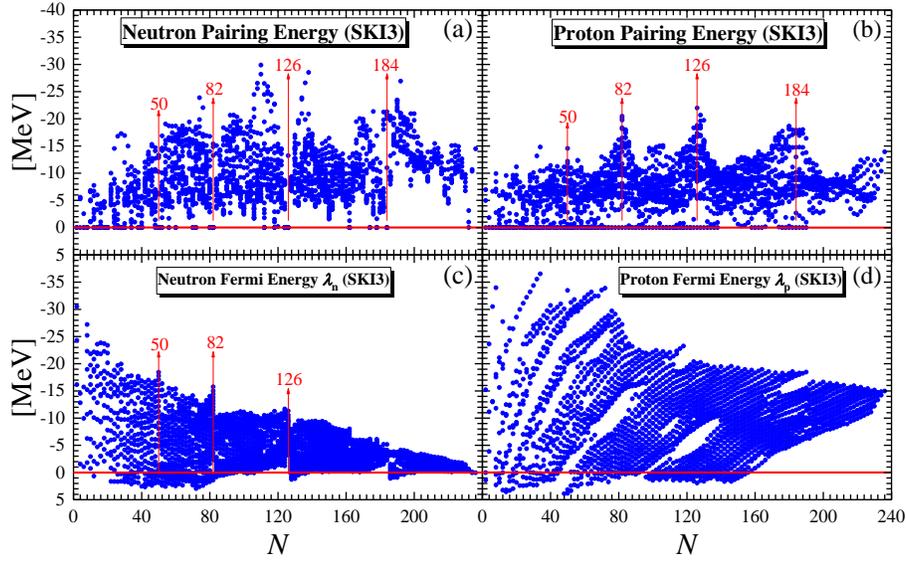

FIG. 8 (Color online) Same as Fig. 7 but as a function of the neutron number *N*.

The calculated results of $E_\Delta^n$, $E_\Delta^p$, $\lambda_n$ and $\lambda_p$ with Skyrme-SKI3 force have been presented as a function of the proton (*Z*) number as shown in Fig. 9(a, b, c, d), respectively. In contrast to the Fig. 8(a), the regions with the magic numbers (*N*=28, 50 and 82) have a large values of $E_\Delta^n$ and appear as a peaks. Remarkably, the regions of those nuclei which located between the magic numbers have a low values of $E_\Delta^n$ as shown in Fig. 9(a), while that regions between the magic numbers (*N*=28, 50 and 82) have a large values of proton pairing energy $E_\Delta^p$ as clear seen in Fig. 9(b). The positive energy values of $\lambda_n$ are concentrated in the nuclei regions with: $4 \leq Z \leq 44, 54 \leq Z \leq 62$ and $80 \leq Z \leq 88$. Whereas the positive energy values of $\lambda_p$ are concentrated in the nuclei regions with $6 \leq Z \leq 110$ as can be seen in Fig. 9(c, d) for $\lambda_n$ and $\lambda_p$, respectively. Both of the $\lambda_n$ and $\lambda_p$ values of the isotopic chain are more regular in the process of decreasing with increasing of the proton number *Z*.

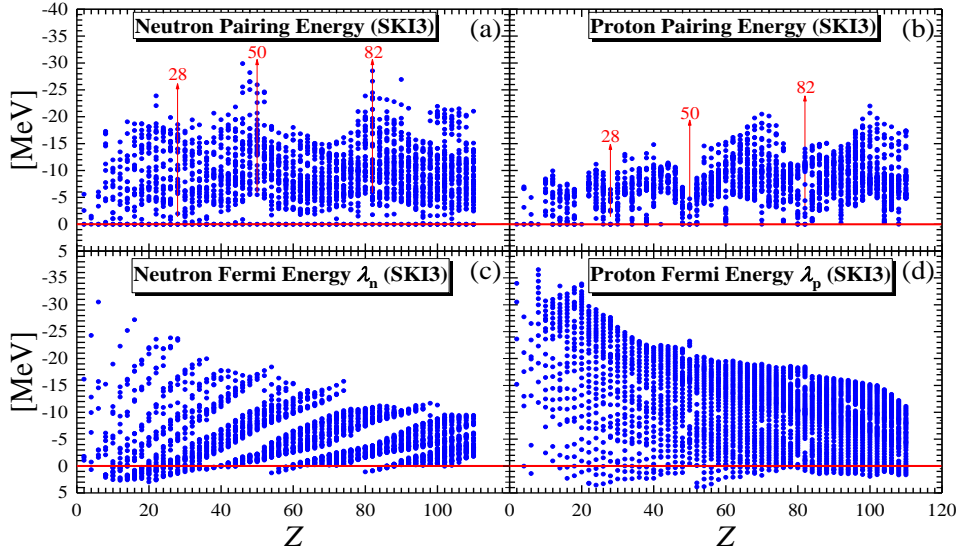

FIG. 9 (Color online) Same as Fig. 8 but as a function of the proton number $Z$.

## 5. CONCLUSIONS

In this work, the ground-state properties for about 2095 even-even nuclei ($2 \leq Z \leq 110$) over a wide range of isotopes ($2 \leq N \leq 236, 4 \leq A \leq 346$) of the whole nuclear chart have been systematically studied by using the Skyrme-SKI3 functional with a mixed volume-surface pairing force interaction in the framework of the Hartree-Fock-Bogoliubov theory. The calculated ground-state properties are including the neutron and proton pairing gap, neutron and proton pairing energy and neutron and proton Fermi energy. The calculated ground-state neutron and proton pairing gap have been compared with the experimental neutron and proton pairing gap of three different formulas $\Delta^{(3)}, \Delta^{(4)}$ and $\Delta^{(5)}$, and also compared with the neutron and proton pairing gap of Lipkin-Nogami model. We conclude that the Skyrme-SKI3 functional with the mixed volume-surface pairing force can be successfully used for describing the neutron and proton pairing gap, neutron and proton pairing energy and neutron and proton Fermi energy of the investigated nuclei, in particularly the neutron-rich nuclei and the exotic nuclei near the neutron drip-line. Moreover, the calculated neutron and proton pairing gap showed good agreement with experimental data of the difference-point formulas $\Delta^{(3)}$, $\Delta^{(4)}$ and $\Delta^{(5)}$, and also with the Lipkin-Nogami model. Finally, the regular agreement regions of the calculated neutron and proton pairing gap with the the difference-point formulas (in particular with $\Delta^{(3)}$ and $\Delta^{(4)}$) were with ($20 \leq Z \leq 110, 20 \leq N \leq 170$), except for some of nuclei (with red color in the chart of nuclei) between these regios which shown deviation in its values, and the regular agreement regions of the calculated neutron and proton pairing gap with the neutron and proton pairing gap of Lipkin-Nogami model were with ($26 \leq Z \leq 110, 26 \leq N \leq 236$), except for some of nuclei (with yellow color in the chart of nuclei) between these regios which shown deviation in its values.

**REFERENCES**

[1] X-YuLiu and C Qi *Comput. Phys. Comm.* **259** 107349 (2021)

[2] M Bender1, K Rutz, P-G Reinhard, and J A Maruhn *Eur. Phys. J. A* **8** 59-75 (2000)

[3] V Thakur, P Kumar, S Thakur, S Thakur, V Kumar, S K Dhiman *Nucl. Phys. A* **1002** 121981 (2020)

[4] T-T Sun, L Qian, C Chen, P Ring, and Z P Li *Phys. Rev. C* **101** 014321 (2020)

[5] N J Abu Awwad, H Abusara, and S Ahmad *Phys. Rev. C* **101** 064322 (2020)

[6] B Dey, S-S Wang, D Pandit, S Bhattacharya, X-G Cao, W-B He, Y-G Ma, N Q Hung, and N D Dang *Phys. Rev. C* **102** 031301(R) (2020)

[7] Z Matheson, S A Giuliani, W Nazarewicz, J Sadhukhan, and N Schunck *Phys. Rev. C* **99** 041304(R) (2019)

[8] P Möller and J. R. Nix *Nucl. Phys. A* **536** 20-60 (1992)

[9] S. Mizutori, J Dobaczewski, G A Lalazissis, W Nazarewicz, and P-G Reinhard

*Phys. Rev. C* **61** 044326 (2000)

[10] P. Ring and P. Schuck *The Nuclear Many-Body Problem* (New York: Springer-Verlag*) First Edit*. M Goldhaber (1980)

[11] M V Stoitsov, J Dobaczewski, W Nazarewicz and P Ring *Comput. Phys. Comm.* **167** 43-63 (2005)

[12] M V Stoitsov, N Schunck, M Kortelainen, N Michel, H Nam, E Olsen, J Sarich, S Wild *Comput. Phys. Comm.* **184** 1592-1604 (2013)

[13] M Bender, P–H Heenen, P-G Reinhard *Rev. Mod. Phys*. **75** 121 (2003)

[14] J Dobaczewski, H Flocard and J Treiner *Nucl. Phys. A* **422** 103-139 (1984)

[15] A. Bulgac *{arXiv: Nuclear Theory} arXiv preprint nucl-th/9907088* (1999).

[16] S A Changizi, C Qi and R Wyss *Nucl. Phys. A* **940** 210-226 (2015)

[17] J Bardeen, L N Cooper and J R Schrieffer *Phys. Rev. C* **108** 1175-1204 (1957)

[18] J A Sheikh and P Ring *Nucl. Phys. A* **665** 71-91 (2000)

[19] N Schunck, M V Stoitsov, W Nazarewicz and N Nikolov *AIP Conf. Proceedings* **1128** 1 (2009)

[20] A H Taqi and M A Hasan *Arab. J. Sci. Eng.* **47** 761-775 (2022)

[21] S A Changizi and C Qi *Phys. Rev. C* **91** 024305 (2015)

[22] S A Changizi and C Qi *Nucl. Phys. A* **951** 97-115 (2016)

[23] P-G Reinhard and H Flocard *Nucl. Phys. A* **584** 467-488 (1995)

[24] P Möller, M R Mumpower, T Kawano and W D Myers *Atom. Data Nucl. Data Tab.* **125** 1-192 (2019)

[25] K Zhang *et al.* (DRHBc Mass Table Collaboration) *Phys. Rev. C* **102** 024314 (2020)

[26] X W Xia, Y Lim, P W Zhao, H Z Liang, X Y Qu, Y Chen, H Liu, L F Zhang, S Q Zhang, Y Kim and J Meng *Atom. Data Nucl. Data Tab.* **121-122** 1-215 (2018)

[27] A H Taqi and M A Hasan *Ukr. J. Phys.* **66(11)** 928-935 (2021)

[28] Y El Bassem and M Oulne *Nucl. Phys. A* **957** 22-32 (2017)

[29] W Satula, J Dobaczewski and W. Nazarewicz *Phys. Rev. Lett.* **81** 3599 (1998)

[30] S J Krieger, P Bonche, H Flocard, P Quentin and M. S. Weiss *Nucl. Phys. A* **517** 275-284 (1990)

[31] S Cwiok, J Dobaczewski, P H Heenen, P Magierski and W Nazarewicz *Nucl. Phys. A* **611** 211-246 (1996)

[32] A. Bohr and B. R. Mottelson *Nuclear Structure Volume I: Single-Particle Motion* (Singapore: World Scientific publishing Co. Pte. Ltd) (1998)

[33] M Bender, K Rutz, P-G Reinhard and J A Maruhn *Eur. Phys. J. A* **8** 59-75 (2000)

[34] M Wang, G Audi, F G Kondev, W J Huang, S Naimi, X Xu *Chin. Phys. C* **41** 030003 (2017)